\documentclass[sigconf]{acmart}
\usepackage{makecell}

\usepackage[strict]{changepage}

\usepackage{framed}

\definecolor{formalshade}{rgb}{0.95,0.95,1}

\AtBeginDocument{%
  \providecommand\BibTeX{{%
    \normalfont B\kern-0.5em{\scshape i\kern-0.25em b}\kern-0.8em\TeX}}}

\copyrightyear{2024}
\acmYear{2024}
\setcopyright{acmlicensed}\acmConference[ICER '24 Vol. 1]{ACM Conference on International Computing Education Research V.1}{August 13--15, 2024}{Melbourne, VIC, Australia}
\acmBooktitle{ACM Conference on International Computing Education Research V.1 (ICER '24 Vol. 1), August 13--15, 2024, Melbourne, VIC, Australia}
\acmDOI{10.1145/3632620.3671098}
\acmISBN{979-8-4007-0475-8/24/08}

\begin{document}

\title[The Appropriation of LLMs in an Undergraduate Programming Course]{Insights from Social Shaping Theory: The Appropriation of Large Language Models in an Undergraduate Programming Course}



\author{Aadarsh Padiyath}
\orcid{0000-0002-4898-3566}
\affiliation{%
  \institution{University of Michigan}
  \city{Ann Arbor}
  \country{USA}}
\email{aadarsh@umich.edu}

\author{Xinying Hou}
\orcid{0000-0002-1182-5839}
\affiliation{%
  \institution{University of Michigan}
  \city{Ann Arbor}
  \country{USA}}
\email{xyhou@umich.edu}

\author{Amy Pang}
\orcid{0009-0000-0077-1487}
\affiliation{%
  \institution{University of Michigan}
  \city{Ann Arbor}
  \country{USA}}
\email{amypang@umich.edu}

\author{Diego Viramontes Vargas}
\orcid{0009-0004-2801-2334}
\affiliation{%
  \institution{University of Michigan}
  \city{Ann Arbor}
  \country{USA}}
\email{dvvargas@umich.edu}

\author{Xingjian Gu}
\orcid{0009-0000-0433-9843}
\affiliation{%
  \institution{University of Michigan}
  \city{Ann Arbor}
  \country{USA}}
\email{xjgu@umich.edu}

\author{Tamara Nelson-Fromm}
\orcid{0000-0001-8197-8366}
\affiliation{%
  \institution{University of Michigan}
  \city{Ann Arbor}
  \country{USA}}
\email{tamaranf@umich.edu}

\author{Zihan Wu}
\orcid{0000-0002-3161-2232}
\affiliation{%
  \institution{University of Michigan}
  \city{Ann Arbor}
  \country{USA}}
\email{ziwu@umich.edu}

\author{Mark Guzdial}
\orcid{0000-0003-4427-9763}
\affiliation{%
  \institution{University of Michigan}
  \city{Ann Arbor}
  \country{USA}}
\email{mjguz@umich.edu}

\author{Barbara Ericson}
\orcid{0000-0001-6881-8341}
\affiliation{%
  \institution{University of Michigan}
  \city{Ann Arbor}
  \country{USA}}
\email{barbarer@umich.edu}

\renewcommand{\shortauthors}{Padiyath, et al.}

\begin{abstract}
  The capability of large language models (LLMs) to generate, debug, and explain code has sparked the interest of researchers and educators in undergraduate programming, with many anticipating their transformative potential in programming education. However, decisions about why and how to use LLMs in programming education may involve more than just the assessment of an LLM's technical capabilities. Using the social shaping of technology theory as a guiding framework, our study explores how students' social perceptions influence their own LLM usage. We then examine the correlation of self-reported LLM usage with students' self-efficacy and midterm performances in an undergraduate programming course. Triangulating data from an anonymous end-of-course student survey (n = 158), a mid-course self-efficacy survey (n=158), student interviews (n = 10), self-reported LLM usage on homework, and midterm performances, we discovered that students' use of LLMs was associated with their expectations for their future careers and their perceptions of peer usage. Additionally, early self-reported LLM usage in our context correlated with lower self-efficacy and lower midterm scores, while students' perceived over-reliance on LLMs, rather than their usage itself, correlated with decreased self-efficacy later in the course.
\end{abstract}




\maketitle

\section{Introduction}
The ability of large language models (LLMs) to solve programming problems has sparked calls for undergraduate programming curriculum changes \cite{prather2023robots, denny2023chat, denny2024computing, farrokhnia2023swot, becker2023programming}. Often, these pro-LLM arguments focus on the capabilities, availability, and usability of LLMs.
This perspective demonstrates \textit{technological determinism} -- implying the capabilities of technology itself should guide the direction of educational reforms \cite{connolly2011beyond, oliver2011technological, selwyn2013distrusting}.
Although LLMs may have significant capabilities, technological determinism overlooks the influential social and cultural dimensions of technology adoption and use~\cite{williams1996social, sancho2020moving}. Decisions about the use of LLMs in education are nuanced and shaped by more than just their capabilities; they also involve educators' and students' preferences, goals, and the broader educational context~\cite{Ko_2024}. History has shown that predicting a technology's success based solely on its capabilities can be misguided. \citeauthor{ames2019charisma} and \citeauthor{reich2020failure} argue that initiatives like `One Laptop per Child' and Massive Open Online Courses (MOOCs) were driven by determinist ethos, and fell short on their goals of democratizing and revolutionizing education at scale \cite{ames2019charisma, reich2020failure}. They suggest this was, in part, due to insufficient consideration of social and individual factors~\cite{reich2020failure, sancho2020moving}. 

Undergraduate programming classrooms are valuable contexts for investigating the social and individual dynamics of LLM use. The diversity of students' goals within these environments \cite{diekman2010seeking, brinkman2016applying, cunningham2022bringing} provides a unique opportunity to study how social perceptions influence students' decisions to use LLMs and how they perceive the tools' impact on their programming education. Understanding that technology is both shaped by and shapes societal practices, our paper employs the \textit{social shaping of technology} (SST) theory \cite{williams1996social} to frame the study. SST stands in contrast to technological determinism by positing an interactive and cyclical, as opposed to sequential or optimal, approach to understanding the development and use of technology~\cite{williams1996social}. According to SST, societal groups are not mere passive consumers of technological innovations; they have agency and substantially shape the ways technology is appropriated. Conversely, technology is not a neutral tool; it exerts a considerable influence on the behaviors of these groups. Using SST theory,
we examine how social factors influence the use of LLMs in an introductory programming course and their impact on student self-perception and learning outcomes. Our research is guided by the following research questions:

\textbf{RQ1}: How do social perceptions influence the usage of Large Language Models in an undergraduate intermediate-level programming course?

\textbf{RQ2}: How does LLM usage relate to programming self-efficacy and midterm scores among undergraduate students in an intermediate-level programming course?

To address these questions, we employ a mixed-methods study of an intermediate-level undergraduate programming course incorporating an anonymous student survey, student interviews, and a regression analysis of midterm performance data with students' self-reported use of LLMs on homework. Our findings aim to provide evidence of the social dynamics surrounding LLM usage in coursework and its implications for the undergraduate programming learning experience.

\section{Literature Review}

\subsection{LLMs in Programming Education}

LLMs are a class of machine learning models that probabilistically generate natural language and code by learning statistical relationships from text documents \cite{bender2021dangers}. Especially popular are OpenAI's generative pre-trained transformer (GPT) models, currently available as ChatGPT\footnote{https://openai.com/blog/chatgpt} (as of March 1, 2024 the free version is known as GPT-3.5, and a \$20 per month more advanced version is known as GPT-4) and Github Copilot\footnote{https://github.com/features/copilot}. The capability of these models to generate, document, and explain code has led researchers to explore their possible uses in computing education \cite{prather2023robots}.

Investigations into LLMs' capabilities highlight opportunities and drawbacks when applying them in programming education \cite{farrokhnia2023swot, prather2023robots}. However, a significant limitation of using these tools for educational purposes is that they were not originally designed with programming instruction in mind \cite{openai2018openai}. As such, several research efforts have focused on adapting the tool for scaffolding programming education, often by restricting its ability to produce code or give `the answer,' while still providing useful aid \cite{liu2024teaching, liffiton2023codehelp, kazemitabaar2024codeaid}, or by using LLM's ability to power more sophisticated scaffolding mechanisms \cite{hou2024codetailor,hou2024integrating}. Yet, even when this limited version is provided, many students may opt to use the original version instead \cite{kazemitabaar2024codeaid}.

In formal educational settings, students are often not required to use LLMs, but some make the personal choice to do so \cite{prather2023robots}. Recent surveys and interview studies discuss possible \textit{technical} aspects (usage patterns \cite{prather2023robots, prather2023weird, forman2023chatgpt, barke2023grounded} or broader affordances and constraints of LLM use \cite{yilmaz2023augmented, forman2023chatgpt, Zastudil2023GenerativeAI}) or \textit{structural} aspects (issues of academic misconduct/unauthorized use \cite{Zastudil2023GenerativeAI, prather2023robots, prather2023weird, lau2023ban, roger2024attitudes}) of LLMs in programming classrooms. However, researchers have yet to investigate the \textit{social} and \textit{cultural} dimensions of students' LLM adoption and use. Our study specifically investigates these social dynamics surrounding LLMs.

Although previous studies have investigated novice use of modified LLM assistants \cite{prather2023weird, forman2023chatgpt}, relatively few have explored the effects of using the LLMs in novice students' programming education \cite{kazemitabaar2023novices, kazemitabaar2023studying}. Early studies \cite{kazemitabaar2023studying, kazemitabaar2023novices} have investigated how self-regulated learners use LLMs for code generation, finding patterns of tool use that can lead to over-reliance such as generating entire solutions to programming problems without engaging with the generated code. However, the effects of LLM usage on students' self-efficacy and their perceptions of learning outcomes are not yet understood. Assessing the impact of LLMs on these factors is essential, as they significantly influence and predict persistence in the field of computing \cite{lewis2011deciding}.

\subsection{Self-efficacy in CS Education}
Self-efficacy refers to individuals' subjective evaluations of their own capabilities to successfully perform an activity \cite{bandura1977social}. Self-efficacy has been found to play an important role in learning as it can impact students' persistence on a learning task, and attitudes when facing learning obstacles \cite{bandura1977self, honicke2016influence, dawson2010effect}. In CS education, students' self-efficacy refers to their perception of competence to complete CS courses and finish programming tasks \cite{zingaro2014peer, tsai2019improving, wiggins2017you, prather2020we, lishinski201928}. Students' CS self-efficacy may vary according to major, gender, and racial and ethnic groups \cite{pirttinen2020study, ojha2024computing}. 
In this work, we compared our classroom with historical patterns of self-efficacy using regression analysis with predictors of gender, under-represented-minority-status, and experience to validate if our sample follows broader self-efficacy trends.

Previous research has demonstrated relationships between students' self-efficacy and help-seeking behavior in CS classrooms. Specifically, \citeauthor{cheong2004motivation} \cite{cheong2004motivation} reported that students' instrumental help-seeking, where they request only the amount of help they need to complete the task individually \cite{martin2021help}, was positively correlated with their self-efficacy beliefs. However, students' executive help-seeking, in which they wish to have the task solved for them \cite{martin2021help}, was negatively correlated with students' self-efficacy \cite{cheong2004motivation}. 
Given that LLM tools can become new valuable help-seeking resources \cite{hou2024effects}, we investigated the relationship between students' self-efficacy and their use of LLM tools when working on programming assignments.

\subsection{Social Shaping of Technology and Society}

Discussions of technology's influence on education are often based on the perspective known as \textit{technological determinism} \cite{oliver2011technological, selwyn2013distrusting}. This viewpoint suggests that the inherent capabilities of a certain technology will inevitably shape human behavior and society \cite{smith1994does}. For example, the belief that the capacity of LLMs to generate code necessitates sweeping changes to programming pedagogy and curricula \cite{denny2024computing, finnie2022robots}. Additionally, the belief that the convenience of quickly produced code will necessarily crush students' motivation to solve problems or learn independently \cite{denny2024computing, lau2023ban, Ko_2023}.

However, others argue that deterministic thinking discounts the vital role that social context plays in the integration of new technologies \cite{williams1996social, Ko_2024}. It also downplays human agency in shaping societal impacts of these technologies \cite{smith1994does, winner2017artifacts, Ko_2024, baym2015personal}. Raw technical capacities are not the sole determiners of LLMs' use and effects. Ultimately, human decision -- instructors', students', and others' -- shapes how we incorporate a technology into our learning practices or reject it. Factors such as students' access, prior experience, goals, and support systems may play a crucial role in determining whether LLMs hinder, alter, or enhance learning processes. Similarly, the integration of LLMs by instructors is strongly influenced by their educational values and beliefs \cite{lau2023ban, winner2017artifacts}. The theory of \textit{social shaping of technology} (SST) argues that societal structures and forces significantly influence how technologies like LLMs are designed, disseminated, and appropriated~\cite{smith1994does, williams1996social, baym2015personal, mackay1992extending}. 

\citeauthor{mackay1992extending} conceptualize social shaping theory as three distinct (but not sequential) spheres: (1) conception, invention, development and design; (2) marketing; and (3) appropriation by users \cite{mackay1992extending}. The first sphere focuses on how technologies are functionally and symbolically encoded by designers to afford certain ends, and how ideology plays a central role in these processes. The second sphere sees marketing as crucial to the social shaping process, as it creates demand for technologies and informs their continued development. The third sphere emphasizes the social appropriation of technologies by users in ways that can differ from the intentions of designers and marketers, as users bring their own interpretations and uses to technologies. In our context, we focus on the third sphere, the appropriation of LLMs in an undergraduate programming course.

\begin{figure*}
    \centering
    \includegraphics[width=1\linewidth]{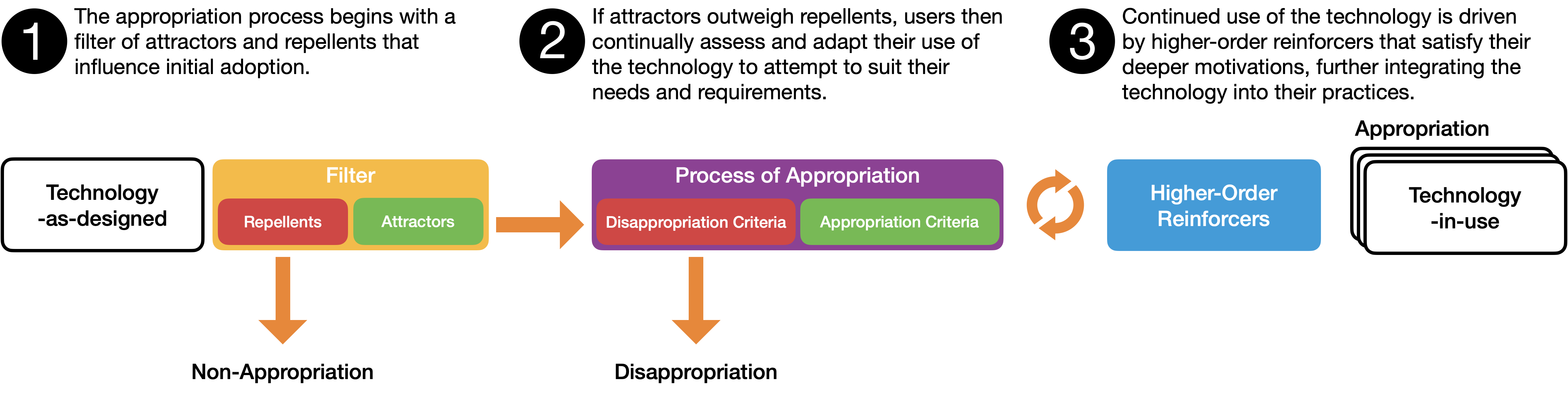}
    \caption{The technology appropriation conceptual model (diagram modified and annotated for clarity from \cite{carroll2002just}).}
    \label{fig:tam}
\end{figure*}

We utilize the technology appropriation model \cite{carroll2001identity, carroll2002just}, as shown in Figure \ref{fig:tam}, to conceptualize our research context and findings. \citeauthor{carroll2002just} describes the transformation of \textit{technology-as-designed} into \textit{technology-in-use} through this process of \textit{appropriation}. This process begins with a filter of \textit{attractors} and \textit{repellents} that determine whether a user will start experimenting with and evaluating the technology, or choose not to adopt it at all (\textit{non-appropriation}). If the user does decide to explore the technology, the process of appropriation involves an assessment of the technology against various \textit{appropriation criteria}. If the technology is a good fit for the user's requirements, it will be appropriated and integrated into the user's practices, becoming \textit{technology-in-use}. However, if negative perceptions of the technology (\textit{disappropriation criteria}) outweigh the benefits, the user may reject or \textit{disappropriate} the technology. Once appropriated, continued use is maintained through \textit{reinforcers}: higher-order drivers that satisfy the user's deeper needs and motivations~\cite{carroll2002just}.

This contrasts with popular models of technology adoption such as the Technology Acceptance Model \cite{davis1989perceived} and Diffusion of Innovation theory \cite{rogers2014diffusion}, which have been criticized for emphasizing the adoption of a static artifact and not accounting for the evolving interaction between users and technology post-adoption \cite{lunceford2009reconsidering, robertson1996role}. In reality, users of technology often have the agency to modify an innovation to better serve their purposes, even after the artifact has been produced \cite{carroll2001identity}, resulting in the technology-in-use differing from the technology-as-designed. For example, the modern bicycle - originally designed for transportation - became a symbol of empowerment in first-wave feminism and is a medium for class expression through accessories \cite{bijker1997bicycles, norcliffe2001ride}.

\section{Methods}

Our study was conducted in the context of an undergraduate Python course. We began with an anonymous survey to understand students' broad attitudes towards and perceptions around LLMs, followed by semi-structured interviews for deeper insights, and ended with a regression analysis of midterm performance with self-reported use of LLMs on homework. We chose this sequence of methods to first understand the landscape and find general patterns we then explored in depth during interviews. Finally, the analysis of midterm performance was conducted post-semester, due to a stipulation of our university's institutional review board. 

\subsection{Context}

\subsubsection{Course Overview}
This study was conducted during the Fall 2023 semester of the 
"Data-Oriented Programming"
course offered by the School of Information (SI) at 
the University of Michigan,
a large public institution in the Midwest United States. The SI Information Science major differs from the College of Engineering's Computer Science (CS) major at 
the University of Michigan
in several ways. The Information Science degree emphasizes the application and management of data and information in real-world contexts, while the CS degree focuses more on the theoretical and technical side of computing and software development. Anecdotally, undergraduate SI students typically envision themselves as future UX designers, data scientists, or \textit{conversational programmers} (students who ``want to communicate effectively about the internals of software, but not write code themselves'' \cite{cunningham2022bringing}); whereas undergraduate CS students typically envision themselves as future software developers. This course was specifically chosen due to its use of Python, a common first language among CS1 courses \cite{becker2019cs1}, and its unique set of students across various academic disciplines. The course aims to develop intermediate Python programming knowledge (variables, loops, strings, functions, basics of object-oriented programming) and introduce students to various elements of data science (input/output, regular expressions, databases, APIs, JSON, BeautifulSoup, Matplotlib). 
During the semester this course was conducted, it was taught by one instructor alongside five graduate teaching assistants (TAs) and nine undergraduate TAs. Every weekday, office hours of at least two hours were scheduled over Zoom and/or in-person, and each session was staffed by a minimum of two TAs. 

To enroll in this course, students must have passed an introductory programming course. As this course satisfies elective requirements across multiple programs, it attracts students from multiple disciplines. While SI majors, for whom the course is a requirement, typically constitute the majority of the class (\textasciitilde 50\%), there is often representation from CS (\textasciitilde 20\%) and Business Administration majors (\textasciitilde 20\%). Notably, CS students typically use this course to develop Python programming expertise, as the CS course sequence is taught using the C++ language.

In this context, an important aspect to consider is that 
the University of Michigan
provided free access to several generative AI models throughout the duration of this study including GPT-4, Llama 2 (7b), GPT-3.5, and Open Journey for all students and instructors through a custom user interface hosted by the university. 

\subsubsection{Course Homework and Midterms}

Each homework assignment (8 total) and project (2 total) asked students to fix bugs and/or create several functions from scratch (see \cite{padiyath2024undergraduate} for an example homework project). 
Note that prior work has found homework instructions in a similar style can be used to generate code from LLMs rather quickly \cite{savelka2023thrilled}.

Two midterm exams were taken through the course's custom interactive Runestone e-book \cite{ericson2020runestone} but proctored in-person, unless a student had learning, illness, or other accommodations. Students were not allowed to use any outside resources (such as the internet and generative AI) while taking the exam other than a single sheet of 8.5 by 11 inch paper with notes (front and back). Midterms were a mix of multiple-choice (including both single-select and multiple-select) questions, fill-in-the-blank questions, Parsons problems (mixed-up lines of code that students must place in order to accomplish a given task) \cite{ericson2023multi}, and write-code questions. Each question was auto-graded by percent correct, and totaled to 200 points. Questions tested concepts covered in lectures and homework assignments.

\subsubsection{Generative AI Policy}

Our university's approach to addressing generative AI is decentralized, with each instructor implementing their own policies. SI is committed to fostering a culture of collaborative and agentic learning. In this course, we extended our definition of collaboration to include the use of generative AI since it can provide simple help at scale, students may use it in their future careers, and because it is freely available from the university. Given the limited understanding of the impact of LLMs on programming education, we encouraged critical engagement with these tools by discussing their limitations and ethical implications in class. To address plagiarism concerns, we required students to cite if and how they used LLMs in their assignments. Students were authorized to use these tools to explain or debug code, but were explicitly dissuaded from using them to generate a complete answer. Additionally, we emphasized the responsibility of students to understand their code by forbidding the use of LLMs on midterm exams. 
The following quote is from the course syllabus:

\begin{quote}
    \textbf{Working with another person on homework is okay, but don't just copy someone else's work. \textit{Each person must contribute an equal amount to the code when you work with other people.}} \textit{You must list who you worked with on your homework in your code.} If we find that your code is identical to another person's and you didn't list that person as someone you worked with then you will receive a 0 on that work. \textit{You can also use generative AI like ChatGPT, but must list that as a thing you worked with and what it helped you with. You must understand the code that you submit. Remember that generative AI can create incorrect code and that you are responsible for the correctness of the submitted code.}
\end{quote}

Note that this is in stark contrast to the CS department policy on collaboration (which includes generative AI). Syllabi for introductory programming courses in the CS department contain the following:

\begin{quote}
    The following are considered Honor Code violations:
\begin{itemize}
    \item Copying or deriving portions of your code from others' solutions. This includes solutions from any source, \textbf{including AI-generated solutions}.

    \item Collaborating to write your code so that your solutions are identifiably similar.
\end{itemize}
\end{quote}

\subsection{Survey}
\subsubsection{Overview}
During the final week of in-person classes for the Fall 2023 semester, we conducted an anonymous survey exploring students' attitudes, perceptions, and use of LLMs. This survey was launched after all instructional sessions but before final project presentations had been completed. It remained open to responses for eleven days, after presentations concluded. Our goal for the survey was to ensure participation from a representative sample of students in the course. To achieve this, we posted an announcement and assignment with a link to the survey on the online class management system (CMS) used by all students. Students were also reminded to complete it during the final project presentation period. In the call to participate, students were made aware of the survey's goal: ``to explore the use (or non-use) of LLMs such as ChatGPT in programming classes." To encourage participation from a range of students, students who completed the survey received 10 class points and were entered into a raffle for a chance to win one of 25 \$20 gift cards.

\subsubsection{Survey Materials}

An initial group of 30 questions were developed. Questions were generated based on our research questions as well as derived from previous surveys on student and instructor perceptions regarding LLM use (such as \cite{prather2023robots}). 

Six think-aloud pilot surveys using a convenience sample of CS first-year undergraduates resulted in the refinement of these questions for clarity and evidence for face validity. This process resulted in the final survey consisting of 25 questions, in addition to five self-efficacy questions (adapted from \cite{midgley2000manual}, a scale with evidence for validity and reliability) and 13 demographic questions
\footnote{The full survey questions are provided here: https://tinyurl.com/LLMSurveyProtocol}.
All questions were mandatory. Beside self-efficacy and demographic questions, the survey was divided into four sections. 

\begin{itemize}
    \item The first section included questions regarding respondents' familiarity with LLM tools. 

    \item The second section included questions about how the respondent uses LLM tools and their perception of their classmates use of LLM tools. 

    \item The third section included questions about why the respondent might use LLM tools. 

    \item The fourth section included questions about any concerns respondents may have with LLM tools.
\end{itemize}

\subsubsection{Data Analysis}

Factors were analyzed using regressions to examine the relationships across and within the respondents. 
Using regression analysis \cite{draper1998applied} allows us to control for the influence of other factors in each regression. When possible, we attempt to control for major and self-efficacy to address omitted variable bias. This mitigates the confounding effect of these variables.
Our research team validated each regression and its assumptions with our university's statistics consultants - experts in regression analysis. Our analysis was completed using the software StataSE.

\subsection{Midterm Performance}

After the completion of the course, we created a dataset marking each of the students' (N=203) citations of LLM use in each homework assignment and project, their midterm scores, the answers to a 20-question 7-point Likert scale self-efficacy survey (adapted from Introductory Programming Self-Efficacy Scale (IPSES) \cite{steinhorst2020revisiting}, a scale with evidence for validity and reliability - with lower scores indicating low programming self-efficacy and higher scores indicating high programming self-efficacy) taken during the third week of the course, and their declared majors.

\subsection{Student Interviews}
\subsubsection{Overview}
Following the survey, we wanted to explore the contexts surrounding students' adoption and the effects of LLM use via semi-structured interviews. These interviews aimed to further explore students' attitudes and experiences regarding their use, or non-use, of AI tools, such as ChatGPT, within the programming classroom. In the first week of the following semester, three weeks after the survey closed, an announcement to recruit interview participants was posted on the course CMS. Students were informed that the objective of the interview was to `understand their experiences and opinions about the use of ChatGPT in 
Data-Oriented Programming.' 
They were assured that all responses would be kept confidential. As an incentive to participate, students were sent a \$20 gift card upon completion of the 30-45 minute interview.

The interview questions were designed to gain a deeper understanding of trends highlighted in the survey analysis. 
Three pilot interviews with a convenience sample of students helped refine questions for clarity and provide evidence for face validity. 
Interviews were conducted online through Zoom videoconferencing. Interviewees were asked prior to the interview for permission to record, and recorded interviews were transcribed by Zoom in full and inspected by the first author for accuracy.

\subsubsection{Data Analysis}
Our interview analysis was guided by a combination of content analysis \cite{harwood2003overview} and structural coding \cite{saldana2021coding}. The initial content analysis determined prevalent sources of influence which contributed to our structural coding framework. Our codebook of student influences is provided in Table \ref{tab:qual_codes}. In applying this codebook, the first and second authors coded the data independently \cite{saldana2021coding}. Then, they jointly reviewed their respective coding to assess agreement, collaboratively discussing and resolving any disagreements. Disagreements were rare, with the main issue being how to label the impact of interviewees' perceptions of their career goals. They resolved this by changing the initial structural code from ``influence from long-term goals" to ``influence from career perceptions" for greater clarity. Then, they categorized participants based on their respective majors and experience levels. Finally, they jointly discussed within-group and between-group comparisons of structural codes \cite{saldana2021coding}.

\begin{table*}[]
    \centering
    \begin{tabular}{m{4.5cm}m{4.5cm}m{5cm}}
    \toprule
        \textbf{Structural Code} & \textbf{Interpretation} & \textbf{Exemplar Quote}\\
        \midrule
       Influence from Career Perception  & Perception regarding their future career that influences how they perceive/use LLMs. & \textit{"I do wanna become like a good software developer. So I can't be too reliant on ChatGPT, but like even like the best software developers, I'm sure, are like putting their errors into ChatGPT."} \smallskip \\
       
       Influence from Peer Perception  & Perception of peers that influence how they perceive/use LLMs. & \textit{"There's not really a way for the college to find out that you like used it, so I'm sure people are still using it."} \smallskip \\
       
       Influence from Self-Perception & A self-perception that influences how they perceive/use LLMs. & \textit{"I think I'm not over-reliant because I write the base or like my draft myself."} \\
       \bottomrule
    \end{tabular}

    \caption{The structural codes pertaining to our research questions used for interview transcript analysis.}
    \label{tab:qual_codes}
\end{table*}

\subsection{Participants}

\subsubsection{Survey Participants}

With the participation of 158 out of a total sample of 203 students (a response rate of 78\%), our sample was relatively representative compared to the class demographics obtained from the University's database.

When analyzing survey data results by self-efficacy, we found Cronbach’s $\alpha$ was internally reliable with $\alpha = .93$. With five questions total, each on a 5-point Likert scale, we averaged each respondent's self-efficacy ratings for the 5 questions to form each student's self-efficacy score (as done in previous research \cite{ericson2023multi, hou2023understanding}). 
To investigate whether our dataset follows broader patterns of self-efficacy, we conducted a regression to predict self-efficacy with the following predictors: major (CS, Information, or Other), gender\footnote{No respondents identified as non-binary.} (Men or Women), whether students are underrepresented minorities (URM: Hispanic/Latine, Black, Native American, Pacific Islander), and college-level programming experience (no experience, 1-2 courses, or 3+ courses). We found correlations in our regression ($R^2$ = .18, F(6, 136) = 4.87, p < .001) that align with prior analyses of programming students' self-efficacy \cite{ojha2024computing}:
A significant predictor of higher self-efficacy among our sample was having taken three or more college-level programming courses as compared to those without any college-level experience ($\beta = .83, t(6, 136) = 2.00, p = .047$), regardless of major, gender, and URM-status.

\subsubsection{Midterm Performance Data}

\begin{figure*}
    \centering
    \includegraphics[width=\linewidth]{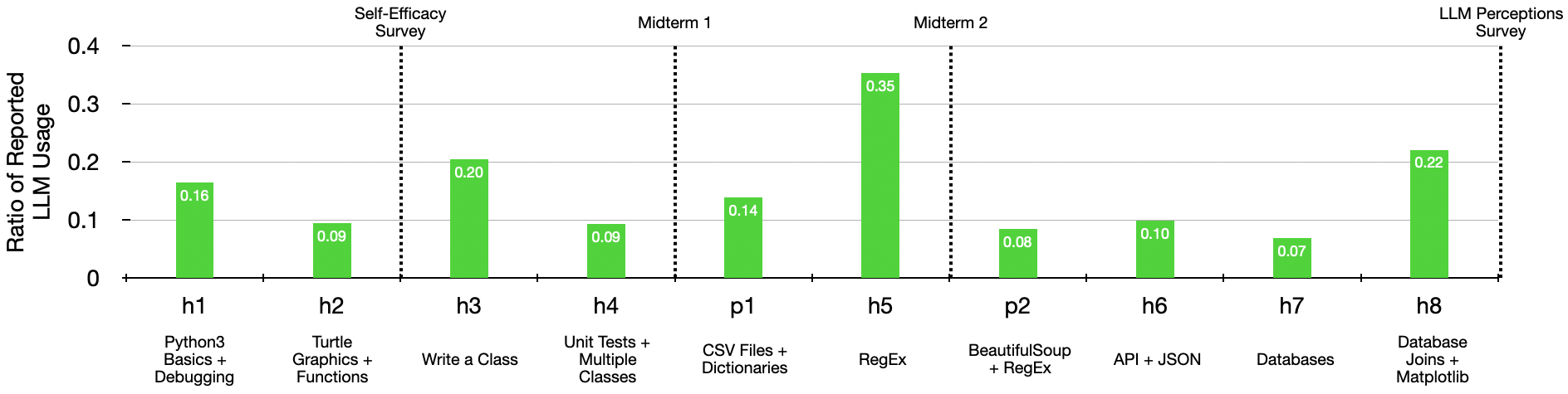}
    \caption{Ratio of students citing LLM usage per each homework assignment/project.}
    \label{fig:llmcitations}
\end{figure*}

The non-anonymous self-efficacy survey was taken in the third week of the course and had a response rate of 78\% (158/203). We found the data was internally reliable with Cronbach’s $\alpha = .96$. With 20 questions total, each on a 7-point Likert scale, we averaged each respondent's self-efficacy ratings over the questions to form each student's self-efficacy score. This was then matched with their midterm scores and LLM use citations to create the midterm performance dataset. LLM Citations throughout the course are visualized in Figure \ref{fig:llmcitations}.

\subsubsection{Interview Participants}

Ten students expressed an interest in participating in the interview study. Their background details are presented in Table \ref{tab:interviewees}. Experience (high or low) was determined from the number of programming classes the student had previously taken (having taken two or more programming classes was categorized as `high' experience).

\begin{table*}[]
    \centering
    \begin{tabular}{cccc}
    \toprule
        \textbf{Participant ID} 
        & \textbf{Major} & \textbf{Experience} & \textbf{Pronouns}\\
         P1 & Business + Information & Low & he/him\\  
         P2 & Business + Psychology & Low & he/him\\ 
         P3 & Business + Information & Low & she/her \\ 
         P4 & Information & Low & she/her\\ 
         P5 & Information & Low & she/her \\ 
         P6 & Information & High & she/her\\ 
         P7 & CS & High & she/her \\ 
         P8 & CS & High & he/him \\ 
         P9 & CS & High & he/him\\ 
         P10 & CS & High & he/him\\ 
         \bottomrule
    \end{tabular}
    \caption{Backgrounds of Interviewees.}
    \label{tab:interviewees}
\end{table*}

\section{Results}

The exploration of our results was driven by our research questions. We present our findings in related sections for clarity. When presenting quotes, ellipses are used for removed text, and square brackets are used for inserting relevant context.

\subsection{Social Perceptions and LLM Adoption}

Through the process of content analysis, we uncovered two social perceptions affecting students' appropriation of LLMs in our context: \textit{perceptions of their future careers}, and \textit{perceptions of their peers' usage}.

\subsubsection{Career Perceptions and LLM Usage}

Survey data relating to career perceptions and to preferences towards LLMs in learning was analyzed using a regression, shown in Table \ref{tab:LLMorProgramming}. The results revealed that students who believed over-reliance on LLM tools would hurt their job prospects in programming careers also tended to prefer learning programming skills themselves, rather than a preference towards learning programming through the use of LLM tools, regardless of major and self-efficacy.

\begin{table*}
    \centering
    \begin{tabular}{p{7.5cm}ccccc}
    \toprule
        "I would rather learn how to use an LLM AI tool like ChatGPT to generate code than learn programming skills myself." & Coefficient & Std. err. & t & P>|t| & Std. $\beta$ \\
         \midrule
        "Over-relying on LLM AI tools like ChatGPT could negatively impact my ability to get a programming job." & -.168 & .0846 & -1.99 & .049 & -.156 \\
        &&&&& \\
        Major (Ref. gp.: Information) &&&&& \\
        Computer Science & -.320 & .241 & -1.33 & .186 & -.110 \\
        Other & -.452 & .231 & -1.96 & .052 & -.157 \\
        &&&&& \\
        Self Efficacy (End of Course) & -.228 & .114 & -1.99 & .048 & -.162 \\
        &&&&& \\
        Constant & 4.038 & .504 & 8.00 & <.001 & \\
        \midrule
        N = 158 &&&&& \\
        F(4, 153) = 4.01 &&&&& \\
        Prob > F = 0.0040 &&&&& \\
        $R^2$ = 0.09 &&&&& \\
         \bottomrule
    \end{tabular}
    \caption{To investigate the relationship between perceptions of LLMs' career impacts and attitudes towards LLMs in programming, we conducted the following regression equation: $LLMorProgramming_s = \beta_0 + \beta_1 ORCareer_s + \beta_2 csMajor_s + \beta_3 otherMajor_s + \beta_3 SE_s + \epsilon_s$. $LLMorProgramming_s$ represents a continuous variable on a 5-point Likert scale regarding students' agreement with the statement, "I would rather learn how to use an LLM AI tool like ChatGPT to generate code than learn programming skills myself." $ORCareer_s$ represents a continuous variable on a 5-point Likert scale regarding students' agreement with the statement, "Over-relying on LLM AI tools like ChatGPT could negatively impact my ability to get a programming job." $csMajor_s$ and $otherMajor_s$ represent dummy variables of students' majors (reference group: information major). Finally, $SE_s$ represents a continuous variable of students' self-efficacy score at the end of the course. $\epsilon_s$ is an error term.}
    \label{tab:LLMorProgramming}
\end{table*}

To gain a more detailed understanding of how career perceptions may influence LLM usage, we asked interviewees when and why they decided to use LLMs in 
Data-Oriented Programming.
Among the explanations interviewees gave, many discussed their future career expectations influencing their approach to using LLMs. 
For example, those targeting software development roles believed LLM skills to be essential, anticipating widespread industry use.

\begin{quote}
    \textit{I think [LLMs are] gonna be used in the field, and if someone wants to go in that field, they should know how to use it like appropriately.} - P7, CS Major, High Experience
\end{quote}

This perspective may have motivated these students to engage with LLMs as a way to prepare for their anticipated professional needs. 

However, those who did not view programming as central to their future work tended toward using LLMs to minimize their engagement with programming concepts they perceived as less relevant to their professional goals. Additionally, these interviewees predict that LLMs will be a mainstay in the future workplace, driving some to use them in class.

\begin{quote}
    \textit{I just kinda wanted to like, get through this class ... I don't really care that much about coding concepts, you know. Like to me, it's just a class. So I guess that might have impacted the way I approach this class as well ... I feel like the stakes for me are not that high as like, a software engineer? You know, like cause I just don't see myself using these concepts that much in the future.} - P4, Information Major, Low Experience
\end{quote}

\begin{quote}
    \textit{When you're at a job at like a company, they're not gonna be like, `don't use ChatGPT.' So like, if it's accessible in the real world, why not in the class?} - P1, Business + Information Majors, Low Experience
\end{quote}

This may suggest that these students may have used LLMs as a tool to bypass learning material they believe is not valuable for their future careers, providing more nuance to the survey finding that career perceptions influence students' approach to learning programming.

Interestingly, even among students targeting software development careers, some reported using LLMs selectively based on their perceptions of the relevance of specific topics to their future work. This may explain the increased reported usage of LLMs for homework \#5, the RegEx-focused homework.

\begin{quote}
    \textit{[In deciding when to use ChatGPT,] I understood what I was going to do, and how I was going to do it. It was like things that I was probably familiar with already. Or maybe it was something I didn't think I needed to know for the future, like [Regular Expressions]. I felt like that wasn't a useful package or anything that I would use in the field. So those things I just threw to ChatGPT ...} - P9, CS Major, High Experience
\end{quote}

This demonstrates that students may use LLMs to bypass learning material they do not think is relevant to their future careers, even if they plan to be a software developer.


\subsubsection{Influence from Perceptions of Peer Use}

When students in the survey rated their own usage of LLMs in course assignments, alongside how often they believed their peers used LLMs, students perceived their peers' LLM usage as significantly higher than was actually reported ($\chi^2$(9, N = 148) = 36.44, p < .0001). In a five-point Likert scale, with low scores suggesting rare usage and high scores suggesting frequent usage, students, on average, rated their own usage as 3.5/5 compared to their perception of their peers usage, on average, as 4.3/5.

To understand why students might overestimate their peers' LLM use, in interviews we asked whether their use of LLMs may differ from their peers' use, and why. We found that most participants believed there was widespread unauthorized LLM use in their other programming courses due to a lack of enforcement. Even though many of the interviewees themselves maintained skepticism about using LLMs against course policy.

\begin{quote}
    \textit{The [CS] classes have, like a huge honor code. You're not allowed to like, put in [copied] code or like use code or anything. So I was just using it to debug my errors ... There's not really a way for the college to find out that you used [LLMs], so I am sure that people are still using it.} - P7, CS Major, High Experience
\end{quote}

\begin{quote}
    \textit{I think that it's cheating, like, if it's not allowed ... If you make [using LLMs] against the rules, people are gonna use it anyway.} - P6, Information Major, High Experience
\end{quote}

Interestingly, in our context where LLM use was permitted, the normalization of these tools changed some students' attitudes and behaviors, even when initially skeptical. As LLM use became more visible within their peer group, students felt more comfortable adopting the tools.

\begin{quote}
    \textit{In the beginning I was pretty apprehensive towards using ChatGPT ... I was scared I'd get reliant, but then, people in my class would be like, `Oh, like I use ChatGPT for this', like, `Oh, it was really helpful with this!' and so that kind of made me like open up to using ChatGPT more, just because, like I saw how normalized it was in class and a lot of times the people I was talking to seemed to have a very good understanding of Python as well, so that kind of took away that initial sense of fear that I had.} - P3, Business + Information Major, Low Experience
\end{quote}

\begin{quote}
    \textit{In the beginning, I was just really scared to use it, because, like, even if they say `Yes, [use LLMs]' I will use it too much, and I'm gonna get caught ... but then someone at my table was like, `It's fine, I'm just gonna use it.' So I'm kind of like, `Oh, yeah, I can use it too.'} - P5, Information Major, Low Experience
\end{quote}

This suggests that the normalization of LLM use by peers can influence students' decisions to adopt these tools, even if they initially have concerns about negative impacts on their learning. 

More experienced students, who had taken their initial programming classes without LLMs, expressed concerns that the widespread use of these tools might be undermining their peers' understanding of programming fundamentals.


\begin{quote}
    \textit{That's the biggest like issue ... is maybe they don't really understand ... how everything works before they just copy and paste.} - P10, CS Major, High Experience
\end{quote}

These concerns suggest the potential risks associated with the influence of perceptions of peer use on LLM adoption, especially for novice programmers who may not have the same foundational skills as their more experienced peers.


\subsection{LLM Usage and Self-Efficacy}

\begin{table*}
    \centering
    \begin{tabular}{lccccc}
    \toprule
        Self Efficacy (third week) & Coefficient & Std. err. & t & P>|t| & Std. $\beta$ \\
         \midrule
        LLM Use on Homework Prior to the Third Week & -.493 & .186 & -2.65 & .009 & -.180 \\
        &&&&& \\
        Major (Ref. gp.: Computer Science) &&&&& \\
        Information & -1.405 & .207 & -6.79 & <.001 & -.612 \\
        Other & -1.494 & .228 & -6.56 & <.001 & -.591 \\
        &&&&& \\
        Constant & 6.25 & .178 & 35.08 & <.001 & \\
        \midrule
        N = 158 &&&&& \\
        F(3, 154) = 21.41 &&&&& \\
        Prob > F < 0.0001 &&&&& \\
        $R^2$ = 0.29 &&&&& \\
         \bottomrule
    \end{tabular}
    \caption{To investigate the relationship between LLM Use and self-efficacy, we conducted the following regression equation: $SE_s = \beta_0 + \beta_1 LLMUsePriorThirdWeek_s + \beta_2 infoMajor_s + \beta_3 otherMajor_s + \epsilon_s$. $SE_s$ represents a continuous variable of students' self-efficacy score at the third week of the course. $LLMUsePriorThirdWeek_s$ is a dummy variable representing if students cited using LLMs on their homework prior to the third week of class. $infoMajor_s$ and $otherMajor_s$ represent dummy variables of students' majors (reference group: CS major). $\epsilon_s$ is an error term.}
    \label{tab:3WeekLLMUseSelfEfficacy}
\end{table*}

\begin{table*}
    \centering
    \begin{tabular}{lccccc}
    \toprule
        Self Efficacy (End of Course) & Coefficient & Std. err. & t & P>|t| & Std. $\beta$ \\
         \midrule
        Overreliant? & -.403 & .144 & -2.79 & .006 & -.233 \\
        &&&&& \\
        Reported LLM Usage & .133 & .065 & 2.04 & .043 & .163\\
        &&&&& \\Major (Ref. gp.: Information) &&&&& \\
        Computer Science & .444 & .172 & 2.58 & .011 & .213 \\
        Other & -.133 & .164 & -.81 & .417 & -.065 \\
        &&&&& \\
        Constant & 3.49 & .240 & 14.52 & <.001 & \\
        \midrule
        N = 153 &&&&& \\
        F(4, 148) = 6.47 &&&&& \\
        Prob > F = 0.0001 &&&&& \\
        $R^2$ = 0.15 &&&&& \\
         \bottomrule
    \end{tabular}
    \caption{To investigate the relationship between students' self-efficacy and their self-perception of over-reliance, we conducted the following regression equation: $SE_s = \beta_0 + \beta_1 OR_s + \beta_2 LLMUsage_s + \beta_3 csMajor_s + \beta_4 otherMajor_s + \epsilon_s$. $SE_s$ represents a continuous variable of students' self-efficacy score at the end of the course. $OR_s$ is a dummy variable representing students' self-perception of their over-reliance on LLMs. $LLMUsage_s$ is a continuous variable representing the number of homework students' reported using LLM assistance. $csMajor_s$ and $otherMajor_s$ represent dummy variables of students' majors (reference group: information major). $\epsilon_s$ is an error term.}
    \label{tab:SEOR}
\end{table*}

Self-efficacy emerged as a factor in evaluating the differing effects of LLM usage on students. We examined the relationship between early LLM usage and self-efficacy, accounting for the students' majors. The analysis included only declared LLM usage in the first two homework assignments, due to the self-efficacy survey being deployed in the third week. Our regression, as shown in Table \ref{tab:3WeekLLMUseSelfEfficacy}, indicated a significant negative association between LLM citations on homework and third-week self-efficacy scores regardless of major: $\beta$ = -.18, t(3, 154) = -2.65, p = .009.
In conducting a regression using the self-efficacy survey from the end of the course, as shown in Table \ref{tab:SEOR}, a binary notion of over-reliance (`Yes, I have felt over-reliant' or `No, I have never felt over-reliant') was significantly negatively associated with self-efficacy ($\beta$ = -.23, t(4, 154) = -2.79, p = .006), regardless of their major and their LLM usage. However, self-reported LLM usage was not significantly correlated with self-efficacy regardless of major, as shown in Table \ref{tab:SEendLLMUsage}. This suggests no significant relationship, or a more complex relationship between self-efficacy and LLM usage at the end of the course.

\begin{table*}
    \centering
    \begin{tabular}{lccccc}
    \toprule
        Self Efficacy (End of Course) & Coefficient & Std. err. & t & P>|t| & Std. $\beta$ \\
         \midrule
         Reported LLM Usage & .035 & .060 & .59 & .556 & .046\\
        &&&&& \\
        Major (Ref. gp.: Information) &&&&& \\
        Computer Science & .600 & .164 & 3.66 & <.001 & .289 \\
        Other & -.101 & .163 & -.62 & .535 & -.049 \\
        &&&&& \\
        Constant & 3.70 & .229 & 16.13 & <.001 & \\
        \midrule
        N = 158 &&&&& \\
        F(3, 154) = 5.52 &&&&& \\
        Prob > F = 0.0013 &&&&& \\
        $R^2$ = 0.10 &&&&& \\
         \bottomrule
    \end{tabular}
    \caption{To investigate the relationship between self-efficacy at the end of the course and students reported LLM usage, we conducted the following regression equation: $SE_s = \beta_0 + \beta_1 LLMUsage_s + \beta_2 csMajor_s + \beta_3 otherMajor_s + \epsilon_s$. $SE_s$ represents a continuous variable of students' self-efficacy score at the end of the course. $LLMUsage_s$ is a continuous variable representing the number of homework students' reported using LLM assistance. $csMajor_s$ and $otherMajor_s$ represent dummy variables of students' majors (reference group: information major). $\epsilon_s$ is an error term.}
    \label{tab:SEendLLMUsage}
\end{table*}

Our regressions cannot determine the \textit{causality} between LLM use and self-efficacy. However, in interviews, when asked about the potential downsides of using LLMs for learning, many students with less programming experience reported a decreased sense of self-confidence when they felt over-reliant on LLMs.

\begin{quote}
    \textit{Sometimes I just, like, copy-paste it and just work on it ... I guess, that impacted the way I understand coding ... I would say that if you, if you give me like a coding assessment right now without access to ChatGPT, I would freak out ... [Using ChatGPT negatively affected] my confidence in my own coding abilities, also understanding how to approach coding problems.} - P4, Information Major, Low Experience
\end{quote}



\subsection{LLM Usage and Learning Outcomes}

\subsubsection{Student Perceptions of LLMs Affecting Learning}

Through a regression analysis as shown in Table \ref{tab:ORLearning}, we found that there was a significantly negative correlation between students' concern about over-relying on LLMs negatively affecting their programming skills and their self-reported usage of LLMs in class, regardless of major and self-efficacy ($\beta$ = -.18, t(4, 153) = -2.26, $p$ = .03). This suggests students who were more concerned about the effect of LLMs on their programming skills reported using LLMs less in class.

\begin{table*}
    \centering
    \begin{tabular}{p{7.5cm}ccccc}
    \toprule
        "Over-relying on ChatGPT could negatively impact my ability to learn programming concepts and skills." & Coefficient & Std. err. & t & P>|t| & Std. $\beta$ \\
         \midrule
        Reported LLM Usage & -.363 & .161 & -2.26 & 0.025 & -.177 \\
        &&&&& \\
        Major (Ref. gp.: Information) &&&&& \\
        Computer Science & .151 & .213 & .71 & .481 & .059 \\
        Other & .299 & .204 & 1.47 & .144 & .119 \\
        &&&&& \\
        Self Efficacy (End of Course) & .203 & .100 & 2.03 & .045 & .165\\
        &&&&& \\
        Constant & 3.12 & .412 & 7.59 & <.001 & \\
        \midrule
        N = 158 &&&&& \\
        F(4, 153) = 3.39 &&&&& \\
        Prob > F = 0.011 &&&&& \\
        $R^2$ = 0.08 &&&&& \\
         \bottomrule
    \end{tabular}
    \caption{To investigate the relationship between perceptions of LLMs' impacts on students' learning and reported usage of LLMs, we conducted the following regression equation: $ORLearning_s = \beta_0 + \beta_1 LLMUsage_s + \beta_2 csMajor_s + \beta_3 otherMajor_s + \beta_3 SE_s + \epsilon_s$. $ORLearning_s$ represents a continuous variable on a 5-point Likert scale regarding students' agreement with the statement, "Over-relying on ChatGPT could negatively impact my ability to learn programming concepts and skills." $LLMUsage_s$ is a continuous variable representing the number of homework students' reported using LLM assistance. $csMajor_s$ and $otherMajor_s$ represent dummy variables of students' majors (reference group: information major). $SE_s$ represents a continuous variable of students' self-efficacy score at the end of the course. $\epsilon_s$ is an error term.}
    \label{tab:ORLearning}
\end{table*}

To understand how students self-reported usage of LLMs correlate with their feelings of over-reliance, we conducted several logistic regressions to predict a binary notion of student over-reliance based on students' self-reported usage of LLMs, regardless of self-efficacy and major. Due to the multiple comparisons problem of conducting repeated regressions, we employ the Bonferroni correction \cite{weisstein2004bonferroni} to adjust our $p$-value threshold to $\alpha$ = .05 / 10 = .005. In Table \ref{tab:lrs}, we present the odds ratio and statistics behind each statistically significant regression. Four behaviors were significantly associated with self-reported over-reliance: generating solutions (e.g. suggesting entire solutions for a coding problem), generating usable code (e.g. suggesting fixes in code for bugs), avoiding asking for help (e.g. asking an LLM instead of asking peers or instructors), and improving programming skills (e.g. using LLMs for generating more efficient code). Six attributes were not significant, suggesting no, a weak, or a more complex relationship: debugging, explaining programming concepts, ideating, resource finding, and reducing stress.

\begin{table*}[]
    \centering
    \begin{tabular}{cccccc}
    \toprule
        Reasons for Using LLMs & LR $\chi^2$(4, N=153) & Psuedo-$R^2$ & Odds Ratio & $p$-value & 95\% Confidence Interval\\
         \midrule
         To Generate Solutions & 31.35 & .16 & 2.2 & .005 & [.98 4.99]\\
         To Generate Usable Code & 32.75 & .17 &3.47& .002 & [1.55 7.75] \\
         To Avoid Asking for Help & 33.55 & .17 &3.7& .002& [1.62 8.44]\\
         To Improve Programming Skills  & 37.94& .19&4.73& < .001 & [2.03 11.02]\\
         \bottomrule
    \end{tabular}
    \caption{To investigate the relationship between students' self-reported over-reliance and motivations for LLM usage, we conducted the following logistic regression equation: $ORLikelihood_s = \beta_0 + \beta_1 motivation_s + \beta_2 csMajor_s + \beta_3 otherMajor_s + \beta_3 SE_s + \epsilon_s$. $ORLikelihood_s$ represents the log likelihood of a students' self-perception of over-reliance. $motivation_s$ is a dummy variable representing students' different motivations for using LLMs. $csMajor_s$ and $otherMajor_s$ represent dummy variables of students' majors (reference group: information major). $SE_s$ represents a continuous variable of students' self-efficacy score at the end of the course. $\epsilon_s$ is an error term.}
    \label{tab:lrs}
\end{table*}

In interviews we asked students for the advantages and disadvantages of using LLMs. The interviewees readily identified their pros and cons based on personal experiences. Some students described using LLMs as a tool for explaining code, helping them understand concepts and error messages. This aligns with the survey finding of explaining programming concepts not being significantly associated with over-reliance.


\begin{quote}
    \textit{It's a teaching tool, right? Or I used it that way. So it sort of taught me how to do it, and walked through the steps for me. It was sort of like seeing a worked problem in a math textbook. So I was able to get a better grip on what RegEx (regular expression) was.} - P6, Information Major, High Experience
\end{quote}

Most interviewees expressed a concern that relying on LLMs for generating code, as well as using LLMs to avoid needing office hours, negatively affected their learning experience which also aligned with the survey correlations.

\begin{quote}
    \textit{To an extent, I think I learned a little less [from using LLMs to generate code.] Not in the beginning, because I knew, like all the concepts in general ... I think I was a little too reliant on SQL stuff ... it doesn't feel as intuitive as I feel it should.} - P7, CS Major, High Experience
\end{quote}

\begin{quote}
    \textit{Personally, I feel like I didn't learn that well in my classes, because I didn't like actively think about the problems for my for my own ... I tend to like just follow the steps that they that [ChatGPT] give me. So in a way, I was put in like a passive role.} - P4, Information Major, Low Experience
\end{quote}

\begin{quote}
    \textit{I think like the beginning, I used it a lot more ... But then, later, I kind of figured out something is just wrong. So I just I kind of use office hours more.} - P5, Information Major, Low Experience
\end{quote}

Students had positive perceptions of their experiences with LLMs when using them to explain concepts or error messages, however, when using it for learning new concepts and generating code, many students noted these as perhaps being less "intuitive" as they expect it should be. Students who had strong foundations in programming from previous classes without LLMs were self-aware of this narrative, as they reported that their foundations were helpful when they later used LLMs for their work.

\begin{quote}
    \textit{Since I had established my foundations in programming, [Copilot] was extremely helpful ... but I'm seeing my friends and my peers, and it's an absolute struggle. Because think of it as learning math, right? If you don't know how to multiply and divide, and you have a program that does that for you way before you learn how to do algebra, you're not gonna be good at algebra. The same way with computer science. If you start using that in your beginner classes, if your foundations are a generative AI program, you are not gonna be able to make it as a software engineer. You're not even gonna be able to make it to like the harder classes in your major.} - P8, CS Major, High Experience
\end{quote}


\subsubsection{LLM Use and Midterm Performance}

We investigated how LLM use, alongside self-efficacy from the third week of the course and academic major (reference group: CS Major), correlated with students' midterm exam performance. The LLM use factor for regressions regarding the first midterm was based on students citing LLM use for homework assignments \#1-4 -- the assignments submitted prior to Midterm \#1. The LLM use factor for the second midterm was was based on students citing LLM use for homework assignments \#1-5 and Project \#1 -- the assignments/project submitted prior to Midterm \#2.
The results are summarized in Table \ref{tab:midterm_performances}. Results indicate a small negative association of LLM use with Midterm \#1 scores and a significant positive association with self-efficacy measured in the third week for both midterms. Meaning, if students used LLMs prior to Midterm \#1, they correlated with scoring slightly less on Midterm \#1, regardless of self-efficacy and major. 

In investigating the influence of students' first midterm performance on students' decision to use LLMs prior to the second midterm, we conducted a paired t-test focused on students who performed poorly (scoring below average, N=77) on the first midterm. The results of the paired t-test were significant, as students who performed poorly on midterm \#1 and used LLMs prior to that exam were significantly less likely to reporting using LLMs prior to the second midterm (t(76) = 2.137, p = .036).

\begin{table*}[]
    \centering
    \begin{tabular}{ccccccccc}
    \toprule
         && \multicolumn{3}{c}{Midterm 1 Performance} && \multicolumn{3}{c}{Midterm 2 Performance} \\
         && Std. $\beta$ & t(4, 153) & $p$-value  && Std. $\beta$ & t(4, 153) & $p$-value \\
         \midrule
         \makecell{LLM Use on Homework\\Prior to the Midterms}&& -.13 & -2.01 & .046 && -.032 & -.43 & .67 \smallskip \\
         \makecell{Third Week Self-efficacy}&&  .37 & 4.51 & < .001   && .23 & 2.71 & .007  \smallskip \\
         Information Major&&  -.30 & -2.96 & .004   && -.30 & -2.69 & .008 \\
         Other Major&& -.15 & -1.51 & .134  && -.18 & -1.60 & .112  \\
         \midrule
         N &&& 158 &&&& 158 \\
         $R^2$ &&& .31 &&&& .16\\
         p-value &&& < .0001 &&&& < .0001\\
         \bottomrule
    \end{tabular}
    \caption{To investigate the relationship between students' midterm performances and reported LLM usage, we conducted the following regression equation: $Midterm_s = \beta_0 + \beta_1 LLMUsage_s + \beta_2 infoMajor_s + \beta_3 otherMajor_s + \beta_3 SE_s + \epsilon_s$. $Midterm_s$ is a continuous variable representing students' midterm scores. $LLMUsage_s$ represents whether students used LLM assistance prior to each midterm. $infoMajor_s$ and $otherMajor_s$ represent dummy variables of students' majors (reference group: CS major). $SE_s$ represents a continuous variable of students' self-efficacy score at the third week of the course. $\epsilon_s$ is an error term.}
    \label{tab:midterm_performances}
\end{table*}

This correlation was also found in discussions with interviewees on the possible influence and impact of LLM use on their midterm performance. Several novice programmer interviewees reported resorting to cramming after under-performing on practice midterms, leading them to recognize their over-reliance on LLMs and start practicing programming more on their own.


\begin{quote}
    \textit{I was studying using ChatGPT, I was doing the practice exam and then putting in the ones I got wrong into ChatGPT and being like, `Why did I get this wrong?' etc. But then I realized, like, I've been using this platform too much to the point where, like, I don't know a lot of the answers ... I was like. Okay, I need to like, you know, learn on my own a little more.} - P2, Business + Psychology Major, Low Experience
\end{quote}


\section{Discussion}

Our findings suggest that students' decisions to appropriate LLMs are not solely driven by LLMs' affordances and capabilities, but are also significantly associated with social factors such as perceptions of peer use, and of the use of LLMs in their future careers. Further, our findings suggest that LLM use might influence users' perceptions of self-efficacy and their learning outcomes. 

\begin{figure*}
    \centering
    \includegraphics[width=1\linewidth]{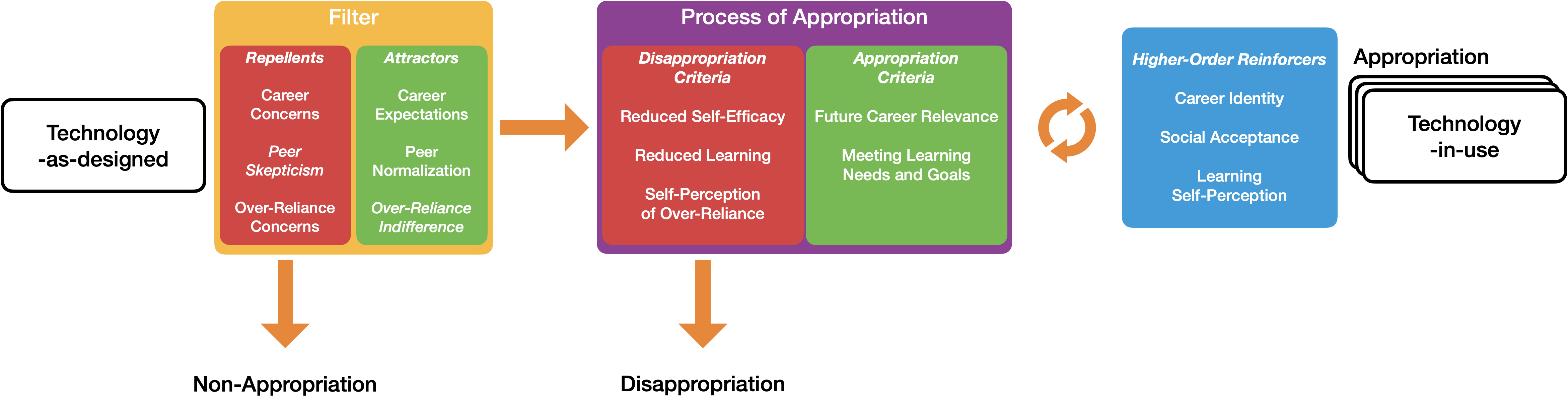}
    \caption{Factors of the Technology Appropriation Model (adapted from \cite{carroll2002just}).}
    \label{fig:tamfilled}
\end{figure*}

Our results encourage us to move beyond a deterministic view of LLMs. We see LLMs as tools whose appropriation is shaped by the social contexts in which they are used. We interpret our key findings in relation to each research question and previous research, as informed by social shaping of technology theory \cite{williams1996social} and the technology appropriation model \cite{carroll2002just}. A summary of our findings is displayed in Figure \ref{fig:tamfilled}. \citeauthor{carroll2002just} suggest attractors and repellents are symmetrical: each attractor is paired with an opposite repellent \cite{carroll2002just}. Thus, we display suggested symmetrical factors, however our data provide no clear examples of factors shown in italicized font.

\subsection{RQ1: How do social perceptions influence the usage of Large Language Models in an undergraduate intermediate-level programming course?}

Our findings suggest student engagement with LLMs is significantly associated with both their perceptions of their career goals, as well as their perception of peers' usage. 
Students who were concerned that over-reliance on LLMs could affect their programming careers told us on the survey that they preferred writing code independently, rather than through LLMs. Their concern aligns with the concept of ``repellents'' in the technology appropriation model, which suggests negative perceptions of a technology can discourage users from adopting it. 
During interviews, almost all students believed LLMs will play a large role in their future careers, or suggested using them to bypass material they anticipate as not having professional value. This selective use of LLMs based on perceived career attributes can be understood through the technology appropriation model's ``appropriation criteria,'' which suggests users assess a technology based on their specific needs and goals when deciding whether and how to adopt it~\cite{carroll2001identity}. Additionally, the tension experienced by students - who are both wary of over-reliance potentially hindering their ability to obtain a programming job, yet anticipate the need for LLMs in their future careers - may be interpreted as a ``higher-order reinforcer''~\cite{carroll2001identity,carroll2002just} -- meaning, the continued use of LLMs is reinforced by students identifying with programming careers, helping them balance their appropriation and disappropriation of LLMs.

Our findings add nuance to \citet{prather2023robots}'s survey and interview findings which suggest students anticipate LLMs playing a large role in their future career. Our study, through the lens of the technology appropriation model, shows that students experience a tension between their career expectations and their perceived necessity of learning to program. This continuously reinforces their decision to either appropriate or disappropriate the tool. This career-based reinforcement may be partly attributed to the broader trend among students to view higher education as a path towards lucrative careers rather than an education towards a greater understanding of the world \cite{Berrett_2022}.

In using LLMs, interviews suggested students exercised agency in determining what they considered essential for learning. Although, as students often do not have authentic industry experience, they can hold misconceptions about their future professional roles and the nature of software engineering \cite{sudal2010analyzing}. For example, interviewees imagined LLMs as a tool freely available to them at their future workplace. However, our participants might be surprised to learn that several companies have imposed bans or restrictions on LLM use (including Apple, Verizon, Amazon, Spotify, Samsung, and Deutsche Bank) \cite{Dugar_2023, DeRose_2023}. Further, the contested legality of LLMs' data collection practices may force companies to destroy LLM products due to copyright infringement \cite{Helms_Krieser_2023}. Alternatively, students may underestimate the role of programming in their future careers, leading them to use LLMs as a shortcut to avoid deeper engagement with programming concepts. As programming becomes an increasingly fundamental skill across career fields \cite{Guzdial_2023, guzdial2023putting}, a solid grasp of programming concepts -- even if not at an expert level -- may be beneficial for students' professional development \cite{Burning_Glass_Technologies_2018}.

Our findings highlight the role of peer perceptions in shaping students' appropriation of LLMs. Survey data showed students may have overestimated their peers' use of LLMs, suggesting a potential mismatch between students' actual usage and their perception of how widely LLMs are used among their classmates. Through the lens of the technology appropriation model, students' perceptions of their peers' LLM usage can be seen as an ``attractor,'' encouraging them to engage with these tools to conform to perceived social norms \cite{carroll2002just}. However, the mismatch may create a sense of social pressure to conform to what is seen as a norm, or otherwise risk a reduced sense of belonging \cite{lewis2016don}. Interview findings support this interpretation, with several participants reporting increased comfort with adopting LLMs after observing their classmates using them, even if they initially had concerns about potential negative impacts on their learning.

Interviewees tended to perceive widespread misuse of LLMs due to a lack of enforceability, even though participants were personally hesitant to use LLMs in violation of class policies. Around the same time we solicited answers for the survey, our university newspaper conducted a similar survey among CS students and found similar results: Most (75\%) of the CS respondents (N=690) reported avoiding using LLM tools on their homework assignments when it was not authorized 
\cite{Shah_Raheja_Sridhar_2024}.
The difference in student LLM usage between these contexts suggests that social acceptance may serve as a higher-order reinforcer for the continued appropriation of LLMs~\cite{carroll2001identity}. As students observe their peers using LLMs, they may prioritize fitting in with their classmates' behavior over their concerns about the potential impact of these tools on their learning. This peer influence may be particularly problematic in a class with such a diverse range of prior programming experience. When less experienced students see their more knowledgeable peers frequently using LLMs, they may assume that adopting these tools will not negatively affect their own learning, even if they lack the same foundational skills as their classmates.

This might demonstrate a limitation of the technology adoption framework, Diffusion of Innovations theory \cite{rogers2014diffusion}. As a component of the framework, \textit{observability} - the extent to which the innovation's benefits are visible to others - does not quite capture this finding. In our study, interviewees attributed adopting LLMs due to their peers' use, rather than adopting due to the benefits, even after considering the possible harmful impact on their learning.

\subsection{RQ2: How does LLM usage relate to programming self-efficacy and midterm scores among undergraduate students in an intermediate-level programming course?}

Social shaping theory posits that technology is not a neutral tool for completing tasks but it can also significantly influence user behavior and beliefs \cite{williams1996social}. Our findings provide evidence for this, as interviewees experienced mixed impact on their programming self-efficacy and self-perception of their learning outcomes when using LLMs. Data from our surveys, interviews, and analysis of midterm performance indicate that reported use of LLMs on course assignments prior to midterm \#1 correlated with decreased self-efficacy and midterm scores. While later use correlated decreased self-efficacy with perceived over-reliance.

One of our more interesting findings is the correlation between early LLM use and decreased self-efficacy, as well as the link between perceived over-reliance on LLMs and lower self-efficacy. In the technology appropriation model, negative perceptions of technology are categorized as ``disappropriation criteria,'' as they may lead users to reject a technology \cite{carroll2002just}. For some students, especially those with less programming experience, the realization of their over-reliance on LLMs and associated decrease in self-efficacy could contribute to disappropriation, prompting them to reduce their usage. However, due to the correlational nature of the survey findings, the directionality of these relationships is still unknown. It is possible that reduced self-efficacy led students to appropriate LLMs. Additionally, interviews with students suggest over-reliance was the factor contributing to their reduced self-efficacy. Larger studies should investigate this potential relationship further.

Given the strong link between self-efficacy and persistence in computing education \cite{lewis2011deciding, beyer2014women, Tek2017ImplicitTA}, decreases in self-confidence are concerning. Over time, this might turn into a belief that they are incapable of understanding the programming concepts without LLM assistance. This relationship is particularly troubling for broadening participation in computing, as perceptions of one's ability to program may be crucial to persistence in computing \cite{APCSP-no-BPC}. 
The possible reduction in self-efficacy likely results from the understanding that more experience with the process of programming fosters greater confidence and persistence in the field \cite{askar2009investigation,ojha2024computing}.

The technology appropriation model recognizes that users can adapt and modify their use of a technology to better suit their needs \cite{carroll2001identity, carroll2002just}. This aligns with the mixed perceptions of LLMs' effects on learning outcomes, with some students reporting positive experiences when using LLMs for explanations and others noting a lack of intuition when relying on them for generating code. Survey data revealed that students concerned about becoming over-reliant on LLMs to learn programming correlated with relying less on LLMs in class. This potentially highlights a ``higher-order reinforcer'' of the need for a perception of learning. Students worried about over-relying on LLMs may disappropriate LLMs to increase their perception of learning. However, this reinforcer may also be problematic, as students who reported using LLMs to improve their programming skills and avoid help-seeking were more likely to feel over-reliant on LLMs.

Students' use of LLMs to generate full solutions to coding problems correlating with over-reliance provides further validation of recent research by \citeauthor{kazemitabaar2023novices}, in which surface-level engagement with the process of programming when fully generating solutions may lead to over-reliance on LLMs \cite{kazemitabaar2023novices}. However, in our context, generating code for debugging may also be related to over-reliance. This may be in contrast to \citeauthor{kazemitabaar2023novices}'s finding that a hybrid approach to using LLM tools -- such as using LLMs for debugging syntax errors -- has a positive effect on students' learning. 
This difference is especially notable as, regardless of major and third-week self-efficacy, reported use of LLMs for homework was negatively correlated with Midterm \#1 performance. However, LLM use was not significantly correlated with Midterm \#2 performance. Through the lens of the technology appropriation model, this difference may be attributed to students realizing that their over-reliance on LLMs was negatively affecting their midterm performance, leading them to disappropriate or modify their use of LLMs. Interviews with students supports this interpretation, as they reported recognizing their over-reliance on LLMs after performing poorly on practice midterms, which prompted some to change their approach.

\subsection{Implications}

A key principle of social shaping theory is that it posits an interactive, cyclical relationship between society and technology, where each continuously influences and reshapes the other \cite{williams1996social, carroll2002just, baym2015personal}. In the context of LLM use in our undergraduate programming class, we begin to see evidence of this interaction. The technology appropriation model helps illustrate this dynamic cycle, emphasizing the role of student agency in the appropriation process.

However, our findings also suggest that students' use of LLMs can have unintended consequences. Some students in our study reported learning setbacks and over-reliance as well as received lower exam scores, potentially leading them to reconsider their dependence on LLMs. This realization may prompt students to modify their engagement with LLMs, initiating another cycle of appropriation based on their updated perceptions and experiences. Additionally, our results indicate that students may not fully grasp the impact of these consequences due to potential misconceptions about their future careers and the nature of programming. Ultimately, the long-term effects of this cyclical relationship remains uncertain and warrants further investigation. As students progress through their education and enter the workforce, early experiences with LLMs may continue to shape their attitudes, skills, and decision-making in complex ways. The ongoing interaction between students and LLMs underscores the importance of providing guidance and support to help students navigate their relationships with LLMs. Longitudinal studies may be necessary to investigate the evolving nature of students' LLM appropriation on their academic, professional, and personal trajectories.

While prior research has found similar concerns among students \cite{forman2023chatgpt, yilmaz2023augmented}, discussions focused on technological interpretations and solutions. For example, in response to finding 45\% of their sample of students being neutral or disagreeing that they understood how to use ChatGPT for academic tasks, the authors advised further education on how best to use ChatGPT \cite{forman2023chatgpt}. When finding students expressing concerns about ChatGPT sometimes providing incorrect answers or finding themselves becoming lazy, the authors recommended providing students with “prompt literacy” skills \cite{yilmaz2023augmented}. Or in interpreting widespread use of an LLM-based tool, the authors assumed this meant that students found the tool beneficial \cite{liffiton2023codehelp, liu2024teaching}. These interpretations and implications fail to account for the social factors influencing students’ adoption and use of LLMs. Students lack of understanding in how to use ChatGPT for learning may stem from their individual learning needs, goals, and contexts, rather than simply lacking technical knowledge. Or, students concerns over incorrect responses and laziness may suggest students would rather disappropriate LLMs than learn how to appropriate it differently. Additionally, widespread use may be driven by social factors such as peer pressure/norms, career expectations, or AI hype and marketing, rather than because the tool itself is beneficial.

This emphasis on technological interpretations and solutions comes from a technological determinism perspective, which implicitly or explicitly presupposes that the adoption and impact of LLMs in programming education are inevitable and desirable. This presumption often fails to account for the agency of students in shaping their own learning experiences and the complex social dynamics that influence the adoption and use of technologies. This perspective assumes that students will passively accept and adapt to LLMs, rather than recognizing students as active participants in the learning process, as they actively negotiate their relationship with LLMs based on their individual needs, goals, and contexts. For example, a perspective coming from technological determinism may argue that "there is little doubt that LLMs ... will have a profound impact on computing education over the coming years" \cite{prather2023robots, Zastudil2023GenerativeAI}, and that "LLMs are here to stay" \cite{prather2024interactions}, urging educators to "embrace these changes or face being left behind" \cite{denny2024computing}. However, our study, through the lens of social shaping theory, finds that the impacts of LLMs are not always desirable, as some students may exercise their agency to reject, negotiate with, or disappropriate LLMs if they do not align with their needs and goals, rather than accepting them as an inevitable part of their learning experience. By acknowledging students' agency and the complex social dynamics that shape the adoption and use of educational technologies, educators and researchers can develop more nuanced and student-centered strategies when discussing LLMs in programming curricula, rather than assuming a one-size-fits-all approach based on technological determinism. Paralleling previous educational technologies that failed to live up to their hype, LLMs' success in education may depend on how well they can be adapted and shaped to fit the social realities and needs of students and classrooms, rather than relying on their capabilities alone. By recognizing this crucial role of social factors in shaping the adoption and impact of LLMs, educators and researchers can develop more realistic strategies for addressing the influence of LLMs in programming curricula.

Finally, this study primarily focused on social influences, while cultural influences could be significant as well. For instance, biases embedded in AI systems could discourage participation from groups who feel alienated or harmed by discriminatory algorithms \cite{benjamin2019race, rankin2020intersectional}. Additionally, the environmental costs of training and running large models may deter engagement from those concerned about climate impacts, which also often fall disproportionately on minoritized communities \cite{bullard1993confronting, bender2021dangers}. Or, people may reject use of LLMs due to the controversial training methods and questionable legality with respect to copyright laws \cite{prather2023robots, bender2021dangers}. It's imperative that future research explores cultural dimensions to understand the nuances influencing LLM engagement.

\subsection{Limitations}

There are some key limitations of this research. First, the context of this study impacts the generalizability of these findings. This study was conducted with a specific demographic makeup of participants, at a single institution, and within one country. While consistent trends emerged from both the interviews and survey responses, these findings may not be universally applicable. Future research should aim to validate our results in different educational settings.

Second, despite a relatively high survey response rates, our results may be subject to selection bias. There is a possibility that those who chose to participate, especially in interviews, had relatively strong opinions about the use or rejection of LLMs. We attempted to mitigate this bias by offering different incentives for participation, but the potential influence of self-selection cannot be entirely ruled out. Additionally, our reliance on self-reported data may introduce response bias. Students may have been reluctant to fully disclose their LLM usage due to perceived shame or stigma associated with these tools. 

Third, it is important to note that the regression analyses in this study are correlational, not causal. While we have identified significant associations between various factors like LLM usage, self-efficacy, and learning outcomes, we cannot definitively establish the directionality of these relationships. There may be omitted variables or endogeneity issues that limit the interpretability of our regression results.

Fourth, the non-anonymous self-efficacy survey was conducted in the third week of the semester, which due to the ephemeral nature of self-efficacy, may be noisy or wholly change throughout the course of the semester \cite{lishinski2022self, lishinski2021all}. This limitation may explain why self-efficacy was not as significant predictor of performance on Midterm \#2 compared to Midterm \#1. However, we stress that this early measurement is crucial for understanding students' initial decisions to use LLMs.

Lastly, our literature review was limited to peer-reviewed research. We recognize the machine learning and LLM research communities frequently utilize non-peer-reviewed repositories such as arXiv.org to publish research. However, we focused solely on traditionally published research due to its increased credibility, as a recent review has raised concerns about the quality of some non-peer-reviewed LLM papers \cite{prather2023robots}. This decision may have led to the omission of relevant contributions or to the replication of existing work.

\section{Conclusion}

This study contributes to the understanding of the social dynamics surrounding the appropriation of large language models (LLMs) in undergraduate programming education by triangulating multiple sources of data within the framing of the technology appropriation model. Our research investigated how social perceptions influenced students' decisions to use LLMs and the perceived and actual impacts of LLM usage on students' programming self-efficacy and midterm performance in an intermediate-level programming course. 

Our findings revealed that students' engagement with LLMs was significantly influenced by their perception of future career norms as well as their perception of peer usage. Additionally, the use of LLMs had mixed impacts on students' self-efficacy and perceived learning outcomes. There was a notable negative correlation between LLM usage and self-efficacy regardless of major and a negative correlation between LLM usage and performance on the first midterm. Our results highlight the complex dynamic between technology and social factors, challenging the notion of technological determinism.
By examining the social perceptions and impacts surrounding LLM usage, we gain a better understanding of how LLMs are appropriated and how they influence students' learning experiences and outcomes.
As LLMs and other AI technologies continue to evolve, it is crucial that we consider the social dynamics that shape their appropriation.

\begin{acks}
Our thanks to the University of Michigan Undergraduate Research Opportunity Program for supporting this project. We are grateful to our reviewers for helpful feedback. We also thank University of Michigan CSCAR service for their invaluable assistance.
\end{acks}

\bibliographystyle{ACM-Reference-Format}
\bibliography{sample-base}

\end{document}